\documentclass{l4dc2026}
\makeatletter
\def\jmlr@heading{}
\makeatother
\jmlrproceedings{}{}
\jmlrvolume{}
\jmlryear{}
\jmlrpages{}

\title[GP-Koopman for Noisy Data]{Inverted Gaussian Process Optimization for Probabilistic Koopman Operator Discovery}
\usepackage{times}
\usepackage{subcaption}
\usepackage{wrapfig}

\coltauthor{%
 \Name{Abhigyan Majumdar} \Email{abmajumdar@ucdavis.edu}\\
 \Name{Navid Mojahed} \Email{nmojahed@ucdavis.edu}\\
 \Name{Shima Nazari} \Email{snazari@ucdavis.edu}\\
 \addr Department of Mechanical \& Aerospace Engineering, University of California Davis%
}

\begin{document}

\maketitle

\begin{abstract}%
Koopman Operator Theory has opened the doors to data-driven learning of globally linear representations of complex nonlinear systems. However, current methodologies for Koopman Operator discovery struggle with uncertainty quantification and the dependency on a finite dictionary of heuristically chosen observable functions. We leverage Gaussian Process Regression (GPR) to learn a probabilistic Koopman linear model from data, while removing the need for heuristic observable specification. We present inverted Gaussian Process optimization based Koopman operator learning (iGPK), an automatic differentiation-based approach to simultaneously learn the observable-operator combination. Our numerical results show that the iGPK method is able to learn complex nonlinearities from simulation data while being resilient to measurement noise in the training data, and consistently encapsulating the ground truth in the predictive distribution.
\end{abstract}


\section{Introduction} 
Koopman Operator Theory presents an elegant approach to obtaining globally linear descriptions of nonlinear dynamical systems in infinite-dimensional function spaces \citep{koopman_hamiltonian_1931}. The spectral reformulation proposed by \citet{mezic_spectral_2005} provided a foundation for data-driven modal decompositions and reinvigorated interest in Koopman Operators in the controls and learning fields. Koopman Operator theory allows for the deployment of rigorous linear system analysis and linear control techniques to nonlinear systems \citep{mezic2021koopman}. Further, such globally valid linear models allow for computationally efficient real-time Model Predictive Control (MPC) \citep{korda_linear_2018}.

Although analytical approaches to Koopman Operator discovery exist \citep{asada_global_2023, mauroy_analytic_2024}, data-driven approaches have become more wide-spread and well studied. The Extended Dynamic Mode Decomposition (eDMD) algorithm \citep{williams_datadriven_2015} and its several variants \citep{abolmasoumi2022robust, colbrook_multiverse_2023} are the most popular methods for obtaining finite approximations of the Koopman Operator from a finite dataset of snapshot pairs \citep{brunton_modern_2021}. However, the eDMD family of methods rely on the user choosing a rich set of observable functions, whose collective expressiveness determines the ability of the algorithm to find a good linear fit for the available dataset. This has led to research into learning the observables, defining the lifted function space, directly from data. In that regard, universal function approximators like Deep Neural Networks (DNNs) \citep{lusch_deep_2018, pan_physics-informed_2020, mallen_deep_2021, nozawa_monte_2024} and Gaussian Processes (GPs) \citep{lian_gaussian_2020, zanini_estimating_2021, bevanda_nonparametric_2024, bevanda_koopman-equivariant_2025} have garnered a lot of interest. Neural network based Koopman models have also been integrated within MPC frameworks, bringing the representation power of DNNs to linear-model-based real-time optimal control \citep{xiao_deep_2020, cisneros_data-driven_2020, yu_efficient_2022, zhang_deep_2024, abtahi2025deep}. Not only do deep Koopman models need more training data, they, similar to eDMD approaches, also struggle to provide uncertainty estimates \citep{frion_koopman_2024}, which are important for safety critical applications, considering that finite-dimensional Koopman predictors are inherently an approximation of the infinite-dimensional linear dynamics. Predictive uncertainty estimates could allow for model-based controllers to quantify and handle the plant-model mismatch explicitly, and reject measurement noise, leading to a better balance between safety and performance \citep{mesbah_stochastic_2016, cairano_stochastic_2017}.

In that regard, blending Gaussian Process Regression (GPR) \citep{rasmussen_gaussian_2008} with Koopman Operator theory holds the promise of uncertainty quantification and non-parametric learning \citep{masuda_application_2019, tsolovikos_dynamic_2024}. The linear nature of the Koopman Operator also addresses the computational challenge of multi-step prediction faced by classical GPR approaches to dynamical system modeling \citep{wang2005gaussian, yogarajah_gaussian_2021}. The Gaussian Process Koopman (GPK) model identification problem has been broken down into a two-stage process - operator identification using traditional subspace methods and then GPR to learn the mapping from the original to the lifted space \citep{lian_learning_2019, lian_gaussian_2020}. \citet{loya_koopman_2023} extended this approach to multi-trajectory data records by leveraging multi-trajectory subspace identification algorithm from \citet{holcomb_subspace_2017}, demonstrating substantial improvements (see Section 6.1 and Fig. 3 from \citet{loya_koopman_2023}). The separation of operator and observable discovery into two distinct and independent stages exposes the subspace GP methods (SSID-GPK) to noise sensitivity and sub-optimal solutions, which is discussed more in Section 3.
\begin{figure}
    \centering
    \includegraphics[width=\linewidth]{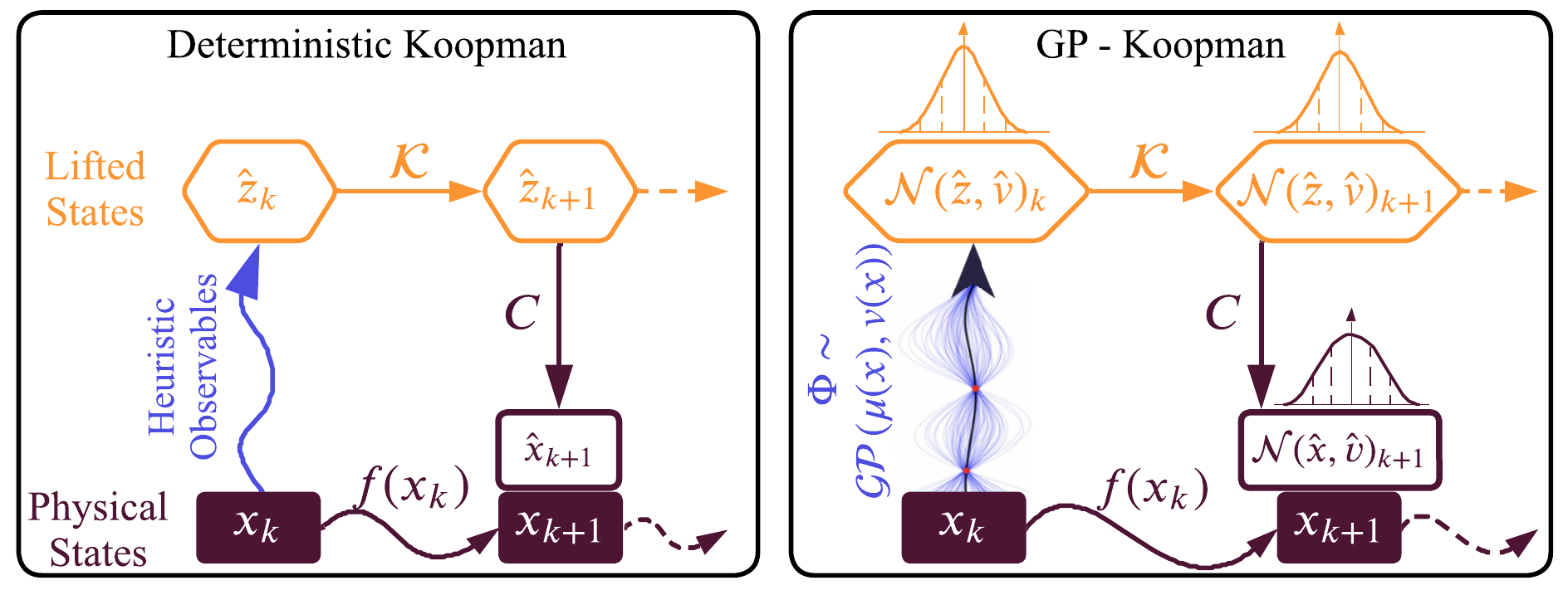}
    \caption{Conceptual Comparison of Deterministic vs Probabilistic GP-based Koopman modeling approaches}
    \label{fig:iGPK Concept}
\end{figure}

While current data-driven Koopman approaches have advanced linear approximation capabilities and easy integration with optimal control, they still have shortcomings such as dependence on manually selecting observable functions, struggle with uncertainty quantification and robustness to noise. To overcome these issues, we propose Inverted Gaussian Process optimization for probabilistic Koopman (iGPK) operator modeling. We simultaneously learn the observables and the Koopman Operator by inverting the standard GPR workflow. The GP training targets are assumed to be \emph{virtual} targets that act as decision variables in an optimization problem. The notable contributions of this work are
\begin{enumerate}
    \item The proposed approach simultaneously learns the lifting space and the Koopman operator by leveraging automatic differentiation and gradient based optimization
    \item The proposed method encapsulates the ground truth within predictive uncertainty bounds
    \item The proposed framework is able to handle observation noise in the training data
\end{enumerate}
We first provide mathematical background for the Koopman operator theory with the eDMD formulation, Gaussian Process Regression (GPR), and the probabilistic GP-Koopman model, as depicted in Fig. \ref{fig:iGPK Concept}. Then, we outline the optimization problem formulation and solution scheme for observable-operator co-discovery. Finally, we present the results of our numerical simulations to compare the predictive performance with other models in the literature, focusing on nonlinear deterministic autonomous systems with observation noise.

\section{Preliminaries}
\subsection{The Koopman Operator}
This section provides a short description of the Koopman Operator with regards to discrete time dynamical systems. For an expansive understanding of the underlying theory and different implementations, readers should refer to \citet{mezic_spectral_2005, mezic2021koopman, brunton_koopman_2016, brunton_modern_2021}.

For the general discrete-time deterministic nonlinear autonomous dynamical system described as
\begin{equation}
    \label{def DynSys Nonlinear}
    x_{k+1} = f\left(x_k\right),\ \forall\ x_k\in\mathbb{X}\subset\mathbb{R}^{n_x},\ f:\mathbb{X}\to\mathbb{X}
\end{equation}
where, $x_k$ is the $n_x$-dimensional state vector at time-step $k$, and $f:\mathbb{X}\to\mathbb{X}$ is a deterministic nonlinear self-map, the Koopman Operator \citep{koopman_hamiltonian_1931}, $\mathcal{K}$, is defined as an infinite-dimensional linear operator on the Hilbert space spanned by the infinite collection of observable functions, $\Phi$, such that
\begin{equation}
    \label{KObasic}
    \Phi \circ f(x_k) = \Phi(x_{k+1}) = \mathcal{K}\Phi(x_k)
\end{equation}
where, $\Phi(x)$ lifts the original system states to a higher dimension and is defined as a collection of individual observable functions $\phi_i(x):\mathbb{R}^{n_x}\to\mathbb{R}$. When we have access to snapshots of data from experiments, i.e., $(x_{k+1}|x_k)_{k=1}^N$, which may or may not be corrupted with observation noise, we can obtain data-driven finite-dimensional approximations of the Koopman Operator. In the Extended Dynamic Mode Decomposition algorithm (eDMD, \citep{williams_datadriven_2015}), $\mathbb{K}_{eDMD}\in\mathbb{R}^{n_z\times n_z}$ is the finite-dimensional approximation of the Koopman Operator, obtained as a least-squares fit of the linear dynamics in the lifted space defined by observable functions $\Phi$. Further, an output linear operator $C_{eDMD}\in\mathbb{R}^{n_x\times n_z}$, is also computed in a least-squares sense to map the lifted states back to the original states.
\begin{equation}
    \label{eDMD}
    \mathbb{K}_{eDMD} = \Phi(X^+)(\Phi(X))^\dagger,\quad C_{eDMD} = X(\Phi(X))^\dagger
\end{equation}
where, $^{\dagger}$ represents the Moore-Penrose left pseudo-inverse of any matrix $A$. When $A$ has full column rank, we define $A^{\dagger}=\left(A^TA\right)^{-1}A^T$. Else, it is defined using the Singular Value Decomposition, $A^{\dagger} = V\Sigma^\dagger U^T$, where $U$ and $V$ are the left and right singular vectors of $A$, and $\Sigma$ is the diagonal matrix of singular values of $A$. $X$ and $X^+$ are the original time-shifted data matrices for $n_T$ trajectories of $N$ time-steps each, obtained from either simulation or experiments
\begin{equation}
    \label{def Dataset}
        X =
        \begin{bmatrix}
            X^{(1)} & ... & X^{(j)} & ... & X^{(n_T)}
        \end{bmatrix}_{n_x\times Nn_T},\quad 
        X^{(j)} =
        \begin{bmatrix}
            x_0^{(j)} & ... & x_k^{(j)} & ... & x_{N-1}^{(j)}
        \end{bmatrix}_{n_x\times N}
\end{equation}
\subsection{Gaussian Processes}
Gaussian processes (GPs) provide a nonparametric Bayesian framework for learning functional mappings with uncertainty quantification. A GP defines a distribution over functions such that for any finite set of inputs, the corresponding function values follow a joint Gaussian distribution \citep{rasmussen_gaussian_2008}. The $i$-th Gaussian process observable (GPO) is expressed as
\begin{equation}
    \label{def GPO}
    \phi_i(x) \sim \mathcal{GP}(\mu_i(x), K_i(x))
\end{equation}
where $\mu_i(x)$ and $K_i(x)$ denote the predictive mean and kernel functions conditioned on training data $\mathcal{D}_i$ and kernel hyperparameters $\theta_i$. The prediction of the $i^{th}$ GPO at any $x$ is a gaussian distribution with mean $\mu_i(x)$ and variance $v_{K,i}(x)$, characterized by the kernel function. These observables provide a natural probabilistic lifting of the state space, capturing both expected values and uncertainty across regions of sparse data \citep{lian_learning_2019, lian_gaussian_2020}.

\subsection{Koopman Operator over Gaussian Process Observable}
We refer to Gaussian Processes modeling the Koopman Observable functions as GPOs. Fig. \ref{fig:iGPK Concept} provides a conceptual comparison between deterministic Koopman models and probabilistic models with GPOs. For the observable function characterized by a GP, such that $\Phi \sim GP(\mu,K)$, the Koopman Operator $\mathcal{K}$ over $\Phi$ is also a GP with $\mathcal{K}\circ\Phi=\mathcal{K}_{\Phi}\sim GP(\mathcal{K}_{\mu},\mathcal{K}_K)$ \citep{lian_gaussian_2020}. We can obtain the mean and covariance functions for the GP Koopman Operator by evaluating the expectation of the lifted states.
\begin{equation}
    \begin{split}
        \mathbb{E}[z_{+}] &= \mathbb{E}[\mathcal{K}_\Phi(x)] = \mathbb{E}_\Phi[\Phi(f(x))]\\
        &= \int_{\mathbb{R}}\Phi(f(x)) p\left(\Phi(f(x))=\epsilon \right)d\epsilon = \mu(f(x))=\mathcal{K}_\mu(x)
    \end{split}
\end{equation}
Similarly, the covariance evolves as
\begin{equation}
    \begin{split}
        cov\left(\mathcal{K}_\Phi\right) &= \mathbb{E}[(\mathcal{K}_\Phi(x)-\mathcal{K}_\mu(x)) (\mathcal{K}_\Phi(x')-\mathcal{K}_\mu(x'))]\\
        &= \mathbb{E}_\Phi[(\Phi(f(x))-\mu(f(x))) (\Phi(f(x'))-\mu(f(x')))\\
        &= K(f(x),f(x')) = \mathcal{K}_{K(x,x')}
    \end{split}
\end{equation}
Both of these are valid because $f:\mathbb{X}\to\mathbb{X}$ is assumed to be a deterministic self-mapping on $\mathbb{X}$ (refer to \citet{lian_gaussian_2020} for a detailed reading). Thus, applying the Koopman Operator to a distribution of lifted states yields another GP distribution, with modified mean and covariance functions, giving us $\mathcal{K}\circ \Phi \sim GP\left(\mathcal{K}_\mu=\mu\circ f, \mathcal{K}_K=K\circ(f\times f) \right)$.

For finite-dimensional approximations applied to system modeling, we define the discrete-time probabilistic Koopman model, with lifted and original state predictions being normal distributions with mean $\hat{z}_k$ and $\hat{x}_k$, and covariance matrices $\hat{\mathcal{V}}_k$ and $\hat{V}_k$ respectively.
\begin{equation}
    \label{def vz prop}
    \hat{z}_{k+1} = \mathbb{K} \hat{z}_k, \quad \hat{\mathcal{V}}_{k+1} = \mathbb{K} \hat{\mathcal{V}}_k \mathbb{K}^T;\ \quad\quad
    \hat{x}_k = C \hat{z}_k, \quad \hat{V}_k = C \hat{\mathcal{V}}_k C^T
\end{equation}
At any time-step $k$, the lifted state is computed using the GPOs defined in Eq. (\ref{def GPO}).
\begin{equation}
    \label{def zk vzk}
    \hat{z}_k = [\mu_{1}(x_k);\dots;\mu_{n_z}(x_k)]_{n_z\times1}, \quad
    \hat{\mathcal{V}}_k= \text{diag}(v_1(x_k),\dots, v_{n_z}(x_k))
\end{equation}
To simplify calculations, we assume each lifted state as being modeled by an individual GPO, $\phi_i(x)$, asserting zero initial covariance, similar to \citet{lian_gaussian_2020, loya_koopman_2023}. However, cross-covariance between different states may appear as a natural consequence of forward propagation via Eq. (\ref{def vz prop}), depending on the structure of the identified Koopman operator, $\mathbb{K}$.

\section{Methodology}
\subsection{Problem Setup}
Any finite-dimensional Koopman Operator-based linear model (of lifted dimensionality $n_z$) of a system (of original state dimensionality $n_x$) consists of three main parts - the linear operator in the lifted observable space ($\mathbb{K}\in\mathbb{R}^{n_z\times n_z}$), the observable functions characterizing that space ($\Phi$), and a mapping from the lifted space to the original state space. While most approaches assume the latter to be a linear mapping ($C\in\mathbb{R}^{n_x\times n_z}$), autoencoder-based Koopman modeling methods learn a separate non-linear mapping function \citep{lusch_deep_2018, pan_physics-informed_2020}. In this study, we consider a linear mapping from the lifted space back to the original state space for simplicity and for preserving the gaussian nature of predictive distributions. \citet{lian_learning_2019} posed the task of learning such a Koopman model as an optimization problem of the form
\begin{equation}
    \label{lian2019 equation}
    \min_{\mathbb{K},\ C,\ \Phi} \| \Phi^+ - \mathbb{K}\Phi \|_F^2 + \| X - C\Phi \|_F^2
\end{equation}
where, $\Phi=\Phi(X)\in\mathbb{R}^{n_z\times n_TN}$ and $\Phi^+=\Phi(X^+)\in\mathbb{R}^{n_z\times n_TN}$ are the lifted data matrices on the available snapshot data pair, $\{X,X^+\in\mathbb{R}^{n_x\times n_TN}\}$ (as defined in Eq. (\ref{def Dataset})), and $\|\cdot\|_F$ denotes the Frobenius norm.

When $\Phi:X\to Z$, is an exact function approximator, the above problem can we rewritten in the form of
\begin{equation}
    \min_{\mathbb{K},\ C,\ Z_0,\ \Phi} \| Z^+ - \mathbb{K}Z \|_F^2 + \left\| X - CZ \right\|_F^2 + \| Z_0 - \Phi(X_0) \|_F^2
\end{equation}
where, $Z_0$ is the set of lifted initial conditions, and the matrices $Z$ and $Z^+$ represent the lifted data matrices, corresponding to $X$ and $X^+$, respectively. In previous works \citep{lian_learning_2019, lian_gaussian_2020, loya_koopman_2023} this problem was solved in two stages - subspace identification \citep{holcomb_subspace_2017} for the first two terms, leading to the solution $(\mathbb{K}^*, C^*, Z_0^*)$; and Gaussian Process Regression \citep{rasmussen_gaussian_2008} for the last term, mapping the original initial conditions ($X_0$) to the lifted initial conditions ($Z_{0}^*$), based on the assumption that GPs are universal function approximators with minimal loss (for details, please refer to Assumption 1 and Footnotes 2 and 3 in \citet{lian_learning_2019}). However, this assumption breaks down when exact fit is intentionally avoided to preserve generalization and especially when the underlying data, $X^+|X$, is corrupted by measurement noise. Thus, in this section, we propose a different optimization-based approach to Eq. (\ref{lian2019 equation}) to obtain a GP-Koopman model.

In Eq. (\ref{lian2019 equation}), we note that the cost varies nonlinearly with the parameters of the function basis $\Phi$, while $\mathbb{K}$ and $C$ appear linearly within the two terms. Following the separable nonlinear least squares approach of \citet{golub2003separable, barligea2023generalized}, Eq. (\ref{lian2019 equation}) can be written as
\begin{equation}
    \label{golub method}
    \begin{split}
        \min_{\Phi}&\left[\min_{\mathbb{K}} \left\| \Phi^+ - \mathbb{K}\Phi \right\|_F^2 + \min_{C}\left\| X - C\Phi \right\|_F^2\right]\\
        = \min_{\Phi}&\left[ \left\| \Phi^+ - \left(\Phi^+\Phi^\dagger\right)\Phi \right\|_F^2 + \left\| X - \left(X\Phi^\dagger \right)\Phi \right\|_F^2 \right]
    \end{split}
\end{equation}
Essentially, the linearly appearing terms, $\mathbb{K}$ and $C$, are eliminated from the cost and the reduced functional is now easier to solve in the parameters of $\Phi$ (see \citet{golub2003separable}). The optimal $\mathbb{K}^*$ and $C^*$ are recovered using the solution to Eq. (\ref{golub method})
\begin{equation}
    \label{optimal K and C}
    \mathbb{K}^* = \Phi^*(X^+)(\Phi^*(X))^\dagger,\quad C^* = X(\Phi^*(X))^\dagger
\end{equation}
The task now is to estimate the optimal functional basis, $\Phi^*$, that maps the original states of the system to the higher-dimensional Hilbert space and minimizes the reduced functional in Eq. (\ref{golub method}). Following previous works \citep{lian_gaussian_2020, loya_koopman_2023}, we assume that each observable is modeled by a separate single-task Gaussian Process, characterized by the choice of kernel function, values of kernel hyperparameters ($\theta_i$ for the $i$-th GPO), and lifted training targets ($Z_i$ for the $i$-th GPO, corresponding to the original initial conditions $X_0$). The challenge however, is in the fact that the training targets for GP Regression, $Z_i$, exist in the lifted, unidentified space, and that the best kernel hyperparameters that capture the mapping between the original state space and the lifted space are also unknown. Thus, for some unknown combination of training targets (in the lifted space) and kernel hyperparameters, we need to minimize
\begin{equation}
    \label{iGPK cost}
        \min_{Z,\ \Theta}\mathcal{L}_1(Z,\Theta) =\min_{Z,\ \Theta}\frac{1}{n_zNn_T}\left[ \left\| \Phi^+ - \left(\Phi^+\Phi^\dagger\right)\Phi \right\|_F^2 + \left\| X - \left(X\Phi^\dagger \right)\Phi \right\|_F^2 \right]
\end{equation}
such that, $Z = [Z_1,\ \dots,\ Z_{n_z}]$, is the set of 'virtual' targets, where each $Z_i\in\mathbb{R}^{n_T,1}$ is the initial condition for the $i$-th lifted dimension, and this is considered a decision variable in our optimization problem. Thus, the training dataset for the $i$-th GPO is defined as $\mathcal{D}_i={(X_0,Z_i)}$, with $X_0 = [x_0^{1},\dots,x_0^{n_T}]\in\mathbb{R}^{n_x\times n_T}$. This is unlike the standard GPR workflow \citep{rasmussen_gaussian_2008}, where target values (for corresponding predictor values) are available as part of a given training dataset. Note that we also normalize the reduced functional by the number of lifted states, time-steps and trajectories, $n_z$, $N$ and $n_T$ respectively, to improve gradient calculation. Further, we define the set of kernel hyperparameters, $\Theta = [\theta_1,\ \dots,\ \theta_{n_z}]$, such that each $\theta_i$ corresponds to the hyperparameters (length-scale, variance and so on) of the covariance kernel, $K$, of the $i$-th GPO. Following \citep{rasmussen_gaussian_2008}, the posterior mean (assuming zero prior mean) of each lifted state, and consequently the lifted matrices ($\Phi$ and $\Phi^+$), can be written as
\begin{align}
    \label{eq:gp_post_mean}
    \mu_{\phi_i|\mathcal{D}_i}(X|Z_i,\theta_i)
        &= \left(K_{XX_0}(\theta_i)\!\bigl(K_{X_0X_0}(\theta_i) + \sigma_i^2 I\bigr)^{-1}Z_i\right)^T,\\
    \Phi(Z,\Theta) &= \left[ \mu_{\phi_1}(X|Z_1,\theta_1);\ \dots; \mu_{\phi_{n_z}}(X|Z_{n_z},\theta_{n_z}) \right]
\end{align}
where $;$ separates different rows of a matrix. Allowing some abuse of notation, we consider the noise assumption, $\sigma_i$, to also be a part of the kernel hyperparameters, $\theta_i$. Note that the transpose operation on the right hand side of Eq. (\ref{eq:gp_post_mean}) has been added to ensure that $\mu_{\phi_i|\mathcal{D}_i}(X|Z_i,\theta_i)$ is a row vector, and that the columns of $\Phi(Z,\Theta)$ represent different data samples (in time and trajectory), while different rows of $\Phi(Z,\Theta)$ represent different state dimensions. This leads to $\Phi(Z,\Theta)$ being a $n_z$-by-$n_TN$ matrix. $\Phi^+(Z,\Theta)=\left[ \mu_{\phi_1}(X^+|Z_1,\theta_1);\ \dots; \mu_{\phi_{n_z}}(X^+|Z_{n_z},\theta_{n_z}) \right]$ is also obtained similarly.
\subsection{Problem Solution}
Although $\mathcal{L}_1$ in Eq. (\ref{iGPK cost}) is a reduced functional, it is still difficult to minimize simultaneously with respect to both $Z$ and $\Theta$. This is because, although $Z$ appears polynomially in $\mathcal{L}_1$, the kernel hyperparameters appear non-linearly in each kernel evaluation, making the gradient computation ($\nabla_\Theta\mathcal{L}_1$) computationally intensive and noisy. Thus, the optimization problem in Eq. (\ref{iGPK cost})  is solved with gradient-descent in 2 stages.

First, we minimize $\mathcal{L}_1$ with respect to $Z$ for randomly initialized $\Theta$. Then, for each GPO, we tune the kernel hyperparameters to maximize the marginal likelihood of observing the optimal virtual training targets, $Z_{i}^{*}$.
\begin{equation}
    \label{cost hp-opt}
    \begin{split}
        \{\min_{\theta_i} &\ \mathcal{L}_2^i\left(\theta_i | Z_i^*,X_0\right),\ \forall\ i\in[1,n_z]\}\ \gets\ \text{arg}\min_{Z} \mathcal{L}_1(Z | \Theta)\\
        \text{s.t.}\ \mathcal{L}^i_2\left(\theta_i | Z_i^*,X_0\right) &=
         \frac{1}{2}\times \left[
                        (Z_{i}^{*})^TK_{\theta_i}(X_0,X_0)Z_{i}^* +
                        \text{log}\left|K_{\theta_i}(X_0,X_0)\right| +
                        n_T\text{log}(2\pi)
                        \right]
    \end{split}
\end{equation}
where $K_{\theta_i}(X_0,X_0)$ is the kernel covariance matrix, evaluated at $X_0$ using the hyperparameters $\theta_i$, and $\text{log}|\cdot|$ represents the log-determinant with base $e$. The second cost function, $\mathcal{L}_2$, is the standard negative log marginal likelihood function for Gaussian Process Regression, the gradient-based optimization of which is widely studied \citep{rasmussen_gaussian_2008, blum2013optimization, chen2022gaussian}. While the virtual target optimization is achieved through Stochastic Gradient Descent (SGD) \citep{sutskever2013importance} to avoid getting stuck in local minima, the the marginal likelihood estimation for $\theta_i$ was achieved by Adam \citep{kingma_adam_2017} as that is more stable for the highly sensitive kernel hyperparameters. The automatic differentiation of the GP posterior mean, and thereby the lifted data matrices with respect to the virtual training targets, ($\nabla_Z\Phi$), was achieved by writing a custom PyTorch-based \citep{paszke2019pytorch} GP-Koopman package. Once the GPOs have been optimized with $\Theta^*$ and $Z^*$ from the gradient-based methods, the Koopman matrices, $\mathbb{K}^*$ and $C^*$, are recovered using Eq. (\ref{optimal K and C}). Algorithm \ref{alg1} illustrates the iGPK approach to learning the optimal $\Phi^*$, $\mathbb{K}^*$ and $C^*$ combination from data.

\begin{algorithm}[h]
\caption{Inverted Gaussian Process Koopman operator learning (iGPK) }\label{alg1}
\BlankLine
\KwIn{System data $\mathcal{D} = \{X_0,X, X^+\}$, number of observables $n_z$, GP prior kernel family $K(\cdot, \cdot|\theta)$}
\KwOut{Optimal Koopman Matrices ($\mathbb{K}^*,\ C^*$) and Optimal GP Observables ($\Phi_{Z^*,\Theta^*}$)}
\BlankLine
\textbf{Initialize:} Randomly initialize virtual targets $Z^{(0)}$ and kernel hyperparameters $\Theta$
\BlankLine

\tcp{Optimize virtual targets $Z$ using SGD}
\For{$g = 0,\dots,G_Z-1$}
    {
    Evaluate cost with forward method $\mathcal{L}_1(Z^{(g)} | \Theta)$ $\gets$ Eq.~(\ref{iGPK cost})\;
    
    Calculate $\nabla_Z \mathcal{L}_1(Z^{(g)} | \Theta)$ $\gets$ Backpropagation\;

    Update $Z^{(g+1)} \gets \text{SGD}(Z^{(g)}, \nabla_Z \mathcal{L}_1)$\;
    }
Record $Z^{(*)}$ as optimal virtual target value\;

\tcp{Kernel hyperparameter optimization for every GPO}
\For{$i=1,\dots,n_z$}
    {
    \For{$g = 0,\dots,G_\Theta-1$}
        {
        Evaluate cost with forward method $\mathcal{L}_2^i(\theta_i^{(g)}|X_0,Z_{i}^{(*)})$ $\gets$ Eq.~(\ref{cost hp-opt})\;
        
        Calculate $\nabla_{\theta_i} \mathcal{L}_2(\theta_i^{(g)}|X_0,Z_{i}^{(*)})$ $\gets$ Backpropagation\;
    
        Update $\theta_i^{(g+1)} \gets \text{Adam}(\theta_i^{(g)}, \nabla_{\theta_i} \mathcal{L}^i_2)$\;
        }
    Record $\theta_i^{(*)}$ as optimal kernel hyperparameter value\;
    }
Compute Koopman Matrices: $\mathbb{K}^*,\ C^*\gets \text{eDMD}\left(X,\ X^+ | \Phi_{Z^*,\Theta^*}\right)$ \;
\BlankLine
\Return{$\mathbb{K}^*,\ C^*,\ \Phi_{Z^*,\Theta^*}$}
\BlankLine
\end{algorithm}

\section{Numerical Results}
In this section, we demonstrate the efficacy of our method by comparing prediction performance with similar methods from the literature. Notably, we compare with eDMD with polynomial and Radial Basis Function (RBF) observables (referred to as the Poly-eDMD and RBF-eDMD approaches, respectively) \citep{williams_datadriven_2015, brunton_modern_2021}, and with our reproduction of the multi-trajectory Subspace Identification (SSID) based GP-Koopman algorithm (referred to as the SSID-GPK) \citep{loya_koopman_2023}. For the RBF-eDMD, we use thin-plate spline type RBF, with the RBF center points determined with K-Means clustering of the original state-space data. For both the systems presented here, we use the Gaussian RBF as the kernel for our iGPK model. For performance characterization, we compute the Normalized Root Mean Square Error (NRMSE) for each trajectory, expressed as a percentage
\begin{equation}
    \%\ \text{NRMSE}^{(j)} = \frac{\sqrt{\frac{1}{N+1}\sum_{k=0}^{N}\left(\|x_{k}^{(j)}-\hat{x}_{k|0}^{(j)}\|^2 \right)}}{\text{max}(x_{k=0\to N}^{(j)}) - \text{min}(x_{k=0\to N}^{(j)})}\times 100\%
\end{equation}
Further, the probabilistic predictions of the SSID-GPK and iGPK models are compared using the Negative Log Predictive Density (NLPD - for details, see \citet{quinonero2005evaluating}) metric, defined as
\begin{equation}
    \text{NLPD} = \frac{1}{N+1}\sum_{k=0}^{N}\frac{1}{2}\left[(x_k-\hat{x}_{k|0})^T\hat{V}_{k|0}^{-1}(x_k-\hat{x}_{k|0}) + \text{log}\left|2\pi\hat{V}_{k|0}\right| \right]
\end{equation}
The first system considered is a scalar discrete-time nonlinear dynamical system with oscillatory behavior, adapted from \citet{zanini_estimating_2021}. The dynamics is governed by
\begin{equation}
    \label{eq zanini system}
    x_{k+1} = -x_k + \frac{3}{1 + x_k^2} + \frac{1}{2}\text{sin}(2x_k)
\end{equation}
\begin{table}[h]
    \centering
    \begin{tabular}{|c||c|c|c|c|}
        \hline
         & Poly-eDMD & RBF-eDMD & SSID-GPK & iGPK \\
         \hline
         Clean Data & $23.1 \pm 16.6$ & $17.9 \pm 19.4$ & $18.6 \pm 14.6$ & $\mathbf{12.2 \pm 12.3}$ \\
         \hline
         Gaussian $5\%$ & $19.8 \pm 16.1$ & $28.2 \pm 25.8$ & $15.4 \pm 12.7$ & $\mathbf{8.0 \pm 9.2}$ \\
         Gaussian $10\%$ & $16.9 \pm 15.4$ & $29.2 \pm 25.4$ & $16.8 \pm 10.4$ & $\mathbf{10.8 \pm 10.4}$ \\
         \hline
         Uniform $5\%$ & $21.9 \pm 16.7$ & $20.4 \pm 18.6$ & $17.8 \pm 12.1$ & $\mathbf{9.8 \pm 12.4}$ \\
         Uniform $10\%$ & $20.1 \pm 17.7$ & $22.7 \pm 20.1$ & $33.2 \pm 15.9$ & $\mathbf{12.0 \pm 13.9}$ \\
         \hline
    \end{tabular}
    \caption{Average Test Set Error (\% NRMSE) for the system in Eq. (\ref{eq zanini system})}
    \label{tab zanini system}
\end{table}
\newpage
\begin{wrapfigure}{r}{0.42\linewidth}
    \vspace{-10pt}
    \centering
    \includegraphics[width=\linewidth]{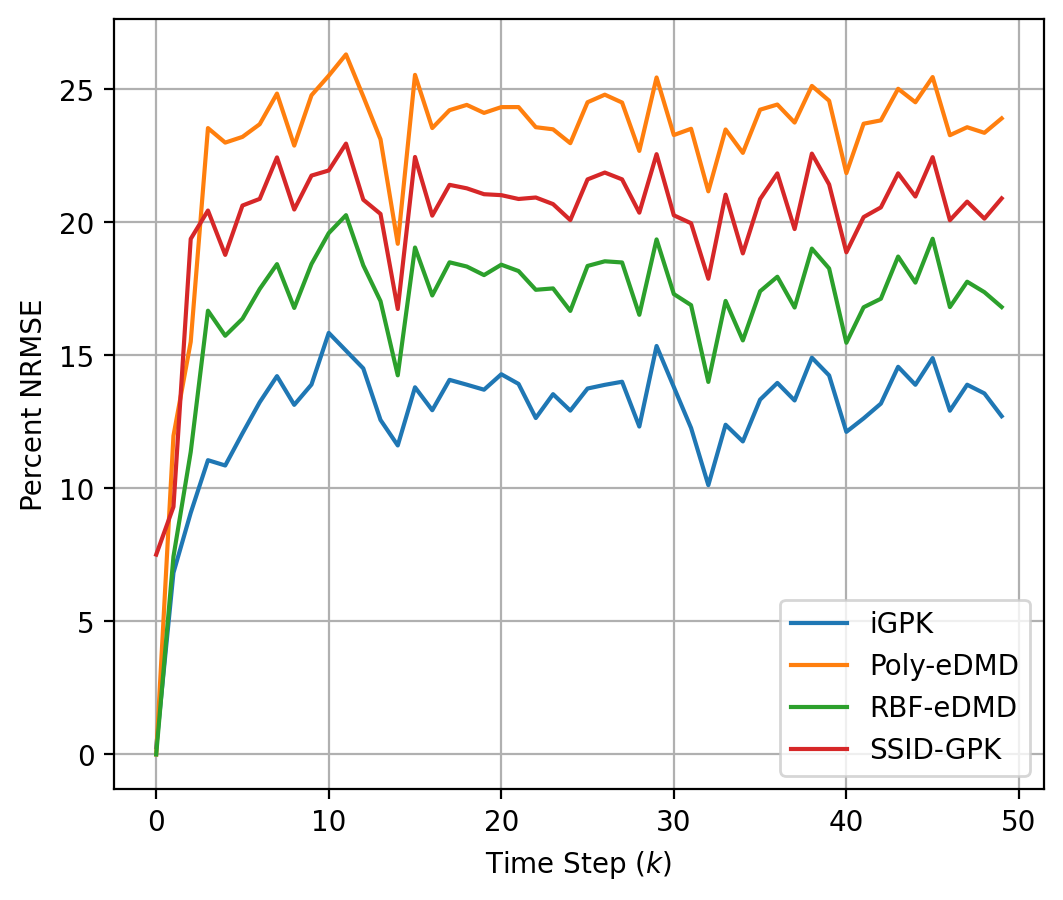}
    \vspace{-8pt}
    \caption{Average test-set cumulative \%-NRMSE (up to step $k$) for the system in Eq.~(\ref{eq zanini system}).}
    \label{fig: sys1 reconstruction}
    \vspace{-5pt}
\end{wrapfigure}
The state transition dataset is generated by randomly sampling 50 initial conditions from a uniform distribution in $[-5,5]$, and simulating for 50 steps. 30 trajectories are used for training, and the rest are reserved for testing. Further, we corrupted the training dataset with zero-mean gaussian noise of 5\% and 10\% intensities. We also considered observation noise from uniform distribution to show model performance in non-gaussian noise scenarios. The model predictions were compared by computing the average (across all the test trajectories) Normalized Root Mean Square Error (NRMSE), expressed as a percentage. Table \ref{tab zanini system} clearly shows how the iGPK model performs better than the legacy methods in all scenario conditions, with a lower mean and standard deviation across trajectories, showcasing not just better, but also more consistent performance. Fig. \ref{fig: sys1 reconstruction} further shows how the iGPK model has lowest the cumulative $\%$-NRMSE across all time-steps for open-loop predictions from initial conditions in the test set.

\begin{figure}[]
    \centering
    \begin{minipage}[t]{0.5\linewidth}
        \centering
        \includegraphics[width=\linewidth]{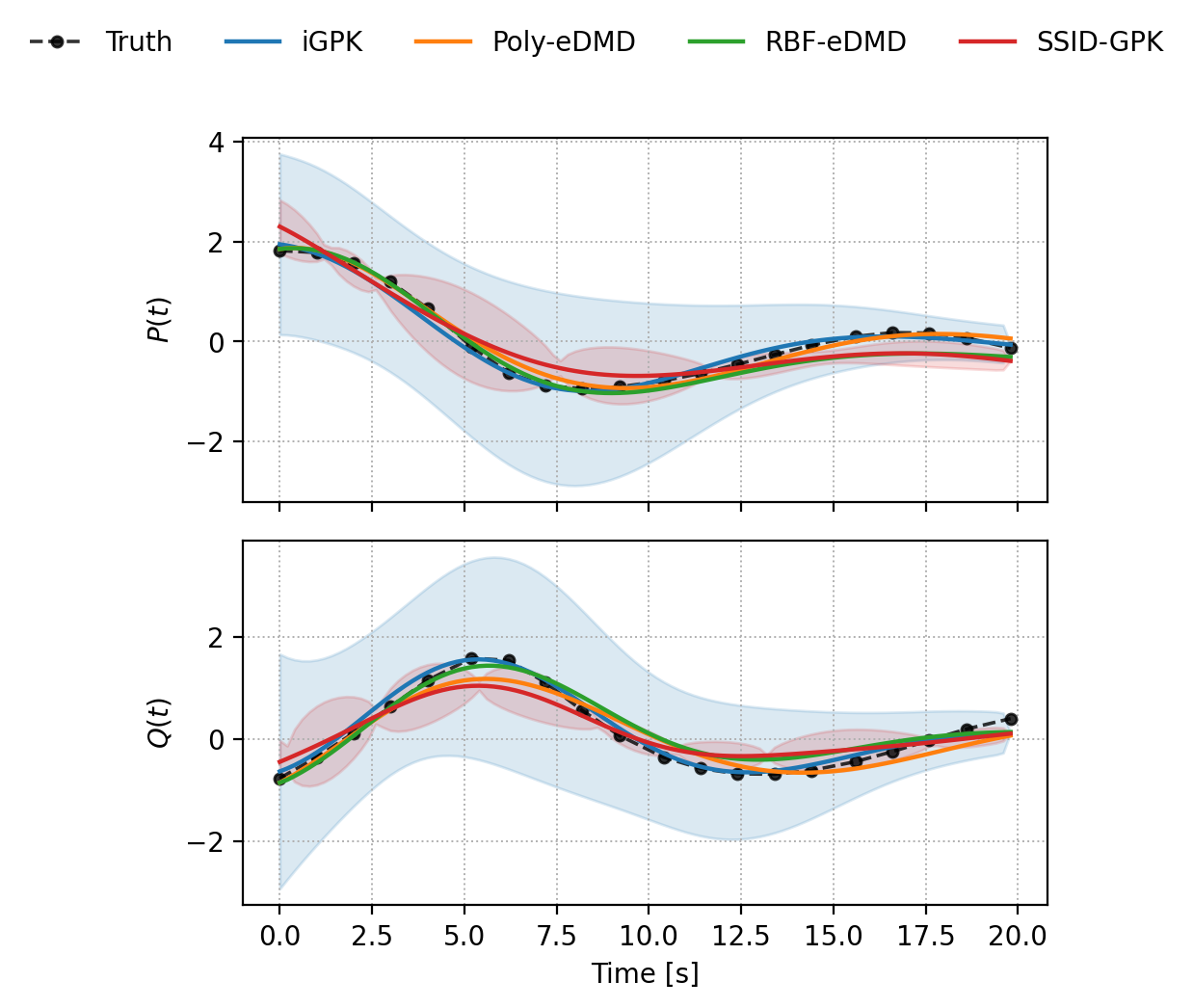}
        \caption*{(A)}
        \label{fig: ipp traj test set}
    \end{minipage}
    \hfill
    \begin{minipage}[t]{0.45\linewidth}
        \centering
        \includegraphics[width=\linewidth]{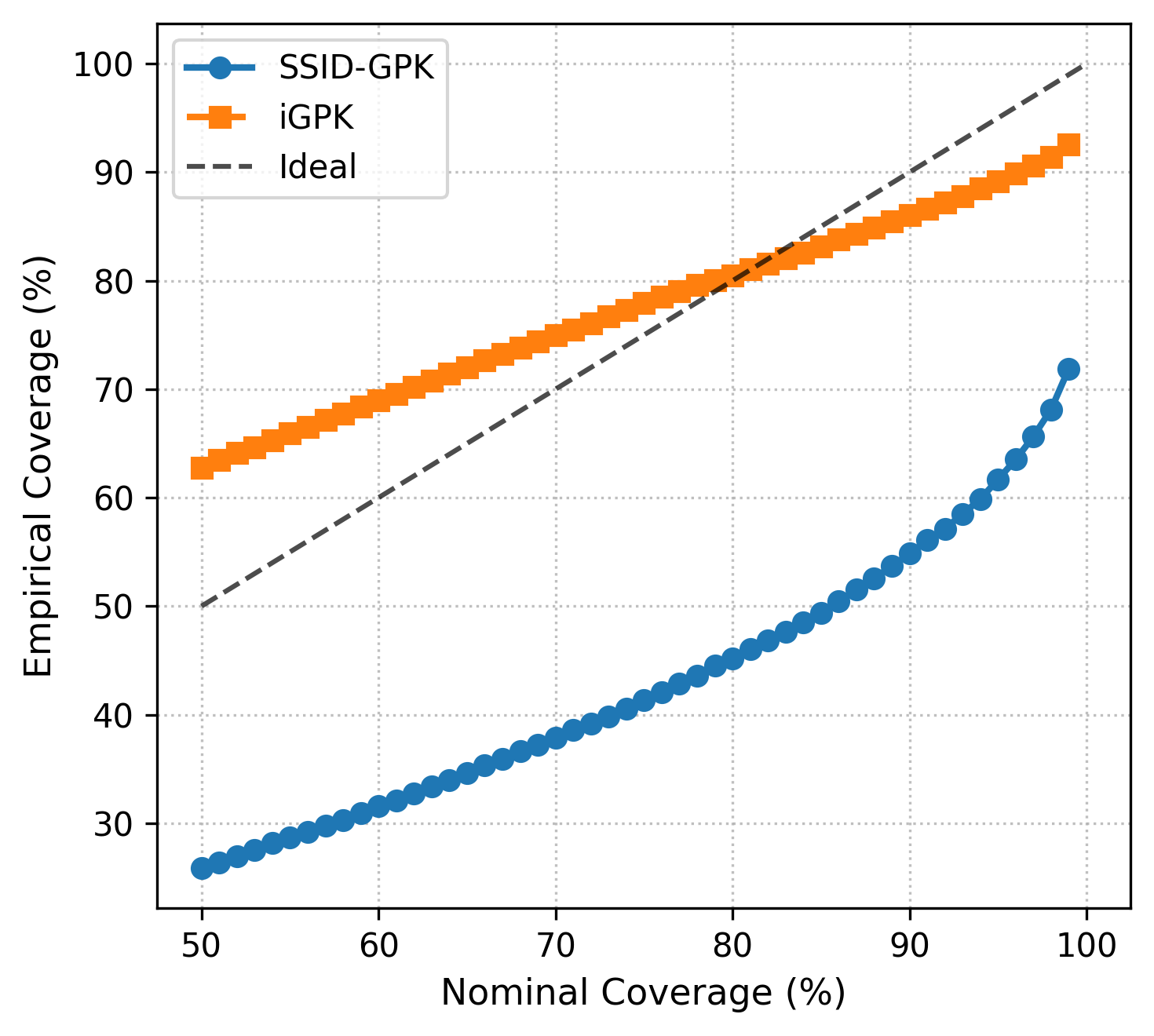}
        \caption*{(B)}
        \label{fig: ipp calibration curve}
    \end{minipage}
    \caption{Results for Predator-Prey system (Eq. (\ref{eq predator-prey})). (A): Open-loop trajectory prediction from test-set initial condition for models trained on data corrupted by $10\%$ Uniform measurement noise. Shaded regions represent the $\pm1\sigma$ predictive uncertainty region; (B) Empirical v/s Nominal Coverage for models trained on data corrupted by $10\%$ zero-mean Gaussian measurement noise}
    \label{figures ipp}
\end{figure}

We also test our method on the Lotka-Volterra Predator-Prey system (with inhibited predation) that describes the interaction between predator and prey populations in an ecological system, considering reproduction of both, and the killing of prey by the predator \citep{lamontagne2008bifurcation, niemann2021data, prakash2023stochastic}.
\begin{equation}
    \label{eq predator-prey}
    \frac{\text{d}P}{\text{d}t} = rP\left(1-\frac{P}{K}\right) - \frac{aP^2}{1+hP^n}Q\ ,\quad
    \frac{\text{d}Q}{\text{d}t} = \eta\frac{aP^2}{1+hP^n}Q - dQ
\end{equation}
Here, $P(t)$ and $Q(t)$, represent the prey and predator populations, respectively, at any given time, $t$. For our studies, we used the parameter values as $r=1,\ K=5,\ a=1,\ h=1,\ n=2,\ \eta=0.5,\ d=0.3$. We sampled 200 initial conditions from a uniform distribution in $[0.1,\ 4]\times[0.1,3]$ and simulated the trajectories for 100 steps with a step-size of $\Delta t=0.2$s, using an RK4 integrator \citep{butcher1996history}. Of these trajectories, 80 were used for training, while the rest were reserved for testing. Fig. \ref{figures ipp}A shows the trajectories predicted by the different models (trained on data corrupted by uniform noise of $10\%$ intensity) for an initial condition from the test set. As we can see, the iGPK model most closely matches the original nonlinear model outputs in both predator and prey populations, while also encapsulating the ground truth within the first standard deviation. Further, we used the large test set to study the coverage performance of the 2 probabilistic models. Fig. \ref{figures ipp}B shows the calibration curves (evaluated on the test set) for the SSID-GPK and iGPK models for training data corrupted with zero mean gaussian noise of $10\%$ intensity. The calibration curve for the iGPK model more closely aligns with the ideal curve, showing better reliability in the predictive distributions for our model \citep{gneiting2007probabilistic}. Table \ref{tab ipp} shows the NLPD for the open-loop predictions on test-set initial conditions for the SSID-GPK and iGPK models. As we can see, the mean NLPD for the iGPK model is consistently below that of the SSID-GPK model, which conveys that the likelihood of observing the ground truth in the iGPK predictive distribution is higher than in the SSID-GPK predictive distribution. More importantly, the standard deviation of the NLPD for the iGPK model is lesser, portraying more consistent performance across all test trajectories.
\begin{table}[!h]
    \centering
    \begin{tabular}{|c||c|c|c|c|}
        \hline
         & iGPK & SSID-GPK \\
         \hline
         Clean Data & $3.18 \pm 0.4$ & $25.12 \pm 60.6$ \\
         \hline
         Gaussian Noise ($10\%$) & $3.32 \pm 2.9$ & $11.78 \pm 21.4$ \\
         Gaussian Noise ($20\%$) & $8.06 \pm 6.7$ & $108.42 \pm 133.2$ \\
         \hline
         Uniform Noise ($10\%$) & $2.76 \pm 0.4$ & $4.57 \pm 6.17$ \\
         Uniform Noise ($20\%$) & $3.88 \pm 2.9$ & $8.64 \pm 14.1$ \\
         \hline
    \end{tabular}
    \caption{NLPD (presented as Mean $\pm$ Standard Deviation across all test set trajectories) for open-loop predictions from the SSID-GPK and iGPK models for the system in Eq. (\ref{eq predator-prey})}
    \label{tab ipp}
\end{table}

\section{Conclusion and Future Work}
In this work, we developed Inverted Gaussian Process optimization for probabilistic Koopman (iGPK) for simultaneous discovery of optimal finite-dimensional Koopman Operator and corresponding GP observables. By treating the GP training targets as virtual targets used as optimization variables, we remove the need for heuristic observable selection. Further, we leverage gradient based optimization and fully differentiable descriptions of GP observables to minimize the linear propagation and mapping losses. Based on our comparisons with other Koopman Operator approaches like eDMD and SSID-based GP-Koopman, we conclude that the proposed iGPK model is superior in capturing nonlinear dynamics from data corrupted with observation noise, especially excelling in quantifying the predictive uncertainty. In the future, we will extend the iGPK method to multi-attractor and non-autonomous systems by utilizing non-stationary and anisotropic kernels, and integrate the probabilistic Koopman model into stochastic linear MPC for fast and robust optimal control of complex nonlinear systems with noisy measurements.
\acks{We used open source software and tooling to aid in refactoring and optimizing the code for this work.}
\bibliography{l4dc2026-sample}

\end{document}